%% file: XAI.tex
\documentclass[runningheads]{llncs}
\usepackage{amsfonts} 
\usepackage{booktabs}
\usepackage{hyperref}
\usepackage[numbers, square]{natbib}
\usepackage[T1]{fontenc}
\usepackage{subfigure}
\usepackage{graphicx}
\usepackage{amsmath}
\usepackage{multirow}
\begin{document}
\title{Here Comes the Explanation: A Shapley Perspective on
Multi-contrast Medical Image Segmentation}
\titlerunning{A Shapley Approach to Multi-contrast Image Segmentation}
%
%

\author{Tianyi Ren\inst{1}\and 
Juampablo Heras Rivera\inst{1} \and
Hitender Oswal\inst{2} \and
Yutong Pan\inst{2} \and
Agamdeep Chopra\inst{1} \and
Jacob Ruzevick\inst{2} \and
Mehmet Kurt\inst{1}
}
\authorrunning{Ren et al.}
%
\institute{Department of Mechanical Engineering, University of Washington,
3900 E Stevens Way NE,
Seattle, WA 98195 \and
Paul G. Allen School of Computer Science, University of Washington,
185 E Stevens Way NE
Seattle, WA 98195 \and 
Department of Neurological Surgery, University of Washington,
1959 NE Pacific Street
Seattle, WA  98195
}
\maketitle              
\begin{abstract}
Deep learning has been successfully applied to medical image segmentation, enabling accurate identification of regions of interest such as organs and lesions. This approach works effectively across diverse datasets, including those with single-image contrast, multi-contrast, and multimodal imaging data. To improve human understanding of these black-box models, there is a growing need for Explainable AI (XAI) techniques for model transparency and accountability. Previous research has primarily focused on post hoc pixel-level explanations, using methods gradient-based and perturbation-based apporaches. These methods rely on gradients or perturbations to explain model predictions. However, these pixel-level explanations often struggle with the complexity inherent in multi-contrast magnetic resonance imaging (MRI) segmentation tasks, and the sparsely distributed explanations have limited clinical relevance. In this study, we propose using contrast-level Shapley values to explain state-of-the-art models trained on standard metrics used in brain tumor segmentation. Our results demonstrate that Shapley analysis provides valuable insights into different models' behavior used for tumor segmentation. We demonstrated a bias for U-Net towards over-weighing T1-contrast and FLAIR, while Swin-UNETR provided a cross-contrast understanding with balanced Shapley distribution.  

\end{abstract}

%
%
%
\input{sec/1_intro}
\input{sec/3_methods}

\input{sec/4_results}
\input{sec/5_Discussion}

\bibliographystyle{splncs04}
\bibliography{sample-ceur}
\appendix
\end{document}

%% file: sec/1_intro.tex
\section{Introduction}
\label{sec:intro}

Segmentation is a fundamental task in medical imaging, involving identifying regions of interest (ROIs) such as organs, lesions, and tissues. By precisely outlining anatomical and pathological structures, segmentation plays a pivotal role in computer-aided diagnosis, ultimately improving diagnostic precision \cite{hesamian2019deep,liu2021review}. Typically, segmentations task are carried out using multi-contrast MRI or multi-modal imaging datasets, due to the necessity of identifying unique microstructural features, such as in gliomas \cite{yan2025clinical}, that are only apparent in some MRI contrasts, but not others.


Many deep learning models, including those used for segmentation, are considered black boxes, offering limited interpretability, resulting in a lack of transparency and accountability \cite{rudin2019stop}. 
Various Explainable AI (XAI) techniques have been developed in the literature \cite{van2022explainable} to tackle this problem, primarily categorized into gradient-based and perturbation-based methods.

\begin{figure}[t]
    \centering
\includegraphics[width=0.7\linewidth, trim=30 60 350 50, clip]{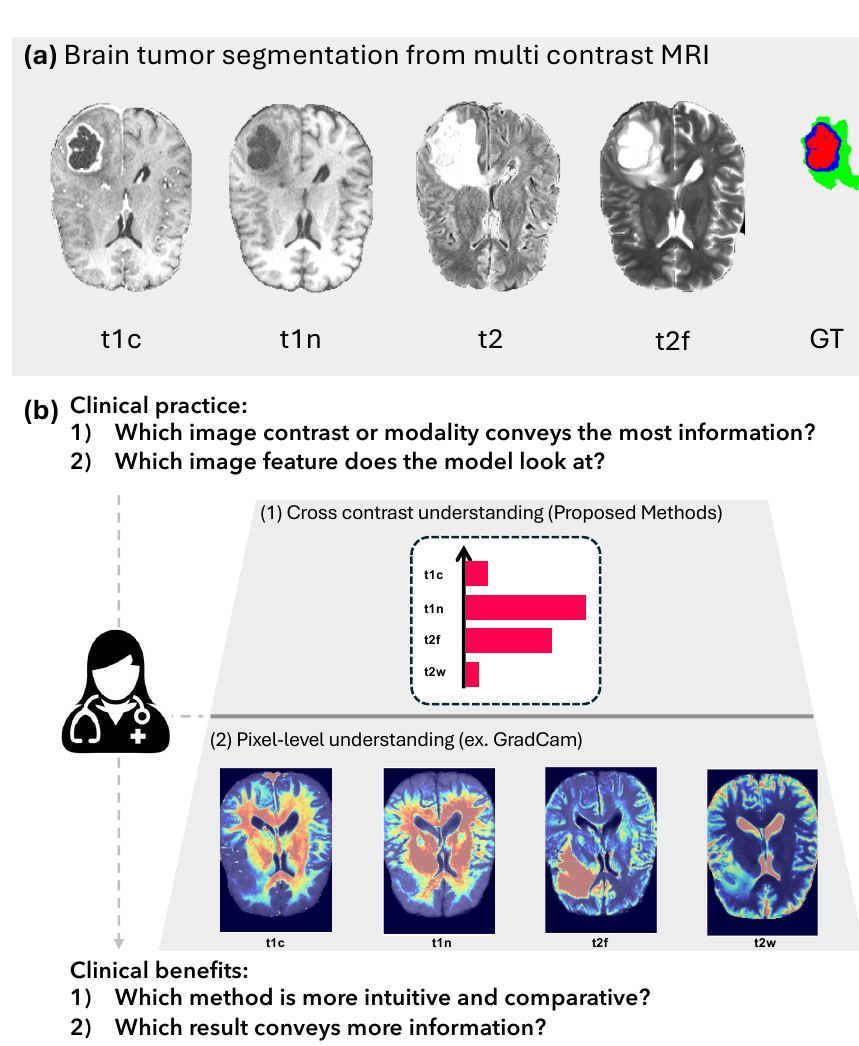}
    \caption{(a) An example of tumor segmentation from multi-contrast MRI. The decision process is not always intuitive because the model does not explain which contrast contributes to the decision, as redundant information can be observed between image contrasts. (b) Our proposed Contrast-level shapley value aims to provide a cross-contrast level explanation which provides a global understanding of the multi-contrast image segmentation. }
    \label{fig:idea}
\end{figure}




Gradient-based techniques, such as saliency maps \cite{saliency} and Grad-CAM \cite{gradcam}, visualize deep learning predictions by identifying influential regions in input data, while perturbation-based approaches (Shapley values \cite{shapley} and LIME \cite{lime}) observe model behavior by systematically perturbing inputs and measuring impact. These methods have been applied successfully to explain the classification problem, however, explaining segmentation still presents significant challenges. There is ongoing debate about whether explanations are necessary for segmentation, as the masks themselves may serve as explanations. Furthermore, there remains uncertainty regarding which components should be explained\textemdash when using gradient-based approaches for models like U-Net, no consensus exists on which layer to target, and in clinical application, which MRI contrasts to explain. Moreover, pixel-level explanations, typically represented as discretized heatmap maps, require further interpretation for grouping analysis \cite{misure}.

  Since in clinical practice radiologists detect lesions by analyzing differences between different MRI contrasts \cite{yan2025clinical}, an explainability framework that reveals deep learning model behavior with regards to how they weigh different MRI contrasts in the segmentation process would be immediately clinically relevant. Therefore, the main objective of this paper is to establish a framework for explaining the contributions of different MRI contrasts in the segmentation process with an application in brain tumor segmentation. This method delivers intuitive quantitative model explanations and enables effective comparisons at multiple levels: between contrasts within a subject (see Figure \ref{fig:3}), and between model architectures for comprehensive model behavior interpretation (see Section \ref{Res-all}). We perform systematic experiments to explain how the state-of-the-art models such as U-Net and Transformer (Swin-UNETR) weigh different MRI contrasts with respect to different evaluation metrics such as Dice and HD95. We conduct statistical analyses to provide an in-depth understanding of how and why different model architectures weigh MRI contrasts differently, even when they achieve similar segmentation performance. In summary, our paper, to the best of our knowledge, is the first study to propose a clinically-relevant explanation framework for brain tumor segmentation in multi-contrast MRI. 



%% file: sec/3_methods.tex
\section{Methods}
\subsection{Dataset and Learning Objectives}\label{data}
The training dataset is sourced from the Brain Tumor Segmentation (BraTS) Challenge 2024 GoAT challenge \cite{baid2021rsna}, consisting of 1,351 subjects. This challenge evaluates algorithmic generalizability across multiple neurological pathologies, including adult glioma, meningioma, and brain metastases. For each subject, four MRI contrasts were given: Native ($t1n$), Post-contrast T1-weighted ($t1c$), T2-weighted ($t2w$), and T2 Fluid Attenuated Inversion Recovery ($t2f$) (Figure \ref{fig:idea}). The ground truth annotations consist of three disjoint classes: Enhancing tumor (ET), Peritumoral edematous tissue (ED), and Necrotic tumor core (NCR). The ET is described by areas that show hyper-intensity in $t1c$ when compared to pre-contrast $t1n$. The appearance of NCR is typically hypo-intense in $t1c$ when compared to $t1n$. Peritumoral edematous/invaded tissue (ED), typically depicted by $t2f$. The detailed preprocessing and training pipeline can be found in our previous research \cite{ren2024optimization,ren2024re}. The model's outputs are used directly for Shapley calculation without post-processing or model ensemble to ensure a fair comparison. 




\subsection{Evaluating Metric for Segmentation}
Evaluating image segmentation involves measuring how effectively an algorithm divides an image into meaningful regions. We used common metrics, including the Dice coefficient and the 95th percentile Hausdorff distance (HD95). 




\vspace{0.3em}
\noindent\textbf{The Dice coefficient (Dice)} \citep{sorensen1948method, sudre2017generalised}, also called the overlap index, is the most widely used metric for validating medical volume segmentations, it measures the overlap between the predicted and ground truth regions. 

\vspace{0.3em}
\noindent\textbf{95th Percentile of Hausdorff distance (HD95)} \cite{huttenlocher1993comparing}. Hausdorff Distance is the maximum surface-to-surface distance between predicted segmentation and ground truth boundary.

\subsection{Model Architectures}
Several state-of-the-art model architectures are tested in this study, including 

\vspace{0.3em}
\noindent\textbf{U-Net} \citep{3DUNet}. The architecture features a symmetric U-shape and consists of two main components: the encoder and the decoder. The encoder, or contracting path, compresses the input volume into a lower-dimensional representation. The decoder, or expansion path, increases the resolution for outputting the segmented map.

\vspace{0.3em}
\noindent\textbf{Seg-Resnet} \citep{myronenko20193d}.  From a U-Net model, a variational auto-encoder branch is added to reconstruct the input image itself to regularize the shared decoder and impose additional constraints on its layers.

\vspace{0.3em}
\noindent\textbf{UNETR} \citep{hatamizadeh2021swin}. UNETR consists
of a transformer encoder that directly utilizes 3D patches and is connected to a CNN-based decoder via skip connection.

\vspace{0.3em}
\noindent\textbf{Swin-UNETR} \citep{hatamizadeh2022unetr}. The segmentation is reframed as a sequence-to-sequence task, where multi-modal input data is embedded into a 1D sequence and processed by a hierarchical Swin Transformer encoder. The encoder extracts multi-scale features at different resolutions using shifted window self-attention, with skip connections linking to an FCNN-based decoder at each resolution.

\subsection{Contrast Level Shapley Value}


Given a training dataset comprised of the pairs $\{ (I, x_0) \}_{i=1}$, where $I \in \mathbb{R}^{4 \times D \times W \times H}$ represents the four 3D-MRI contrast as a multi-channel input, $x_0 \in \mathbb{R}^{3 \times D \times W \times H}$ represents the associated one-hot encoded segmentation mask, with 3 tumor labels: ED, NCR, and ET as described in Section \ref{data}. The deep learning models ($\omega$) were trained to predict the tumor labels $\hat{x}_0$ given the input $I$:
\begin{equation}\label{eq:eq1}
\hat{x}_0 = \omega(I).
\end{equation}



Derived from the Shapley value \cite{shapley}. The Contrast level Shapley value \( \phi_i(\mathit{M}) \) was then evaluated with respect to each specific metric (M) by:

\begin{equation}\label{eq:eq3}
\phi_i(\mathit{M}) = \sum_{S \subseteq N \setminus \{i\}} \frac{|S|! (|N| - |S| - 1)!}{|N|!} \left( \mathit{M}(S \cup \{i\}) - \mathit{M}(S) \right)
\end{equation}
where \( N \) is the set of all of MRI contrasts; \( |N| \) is the total number of contrasts; \( S \)  is a subset of MRI contrasts excluding certain contrast  \( i \) (\( S \subseteq N \setminus \{i\} \)) and  \( |S| \) is the number of contrasts in \( S \); \( \mathit{M}(S) \) is the target metric evaluated on the subset  \( S \). The implementation was developed in PyTorch, allowing user-customized input features for Shapley value calculation. Furthermore, the framework supports Shapley value computation across various medical image segmentation metrics, not limited to Dice and HD95.

\subsection{Statistical test}\label{stats}



The contrast-level Shapley values are examined to assess whether observed differences across folds or between models are statistically significant. Specifically, statistical tests are employed to evaluate differences in group means (central tendency) and variances (dispersion). 


\vspace{0.3em}
\noindent \textbf{Test for equal variance.} Levene's test \cite{levene, devore2012modern, james2023introduction} is applied to assess homogeneity of variance even when the normality assumption cannot be guaranteed \cite{garcia2009study, sun2024spatiotemporal}.





\vspace{0.3em}
\noindent \textbf{Test for equal mean.} If the normality assumption cannot be guaranteed, the Kruskal-Wallis test \cite{theodorsson1986kruskal, devore2012modern} is used instead of ANOVA \cite{fisher1970statistical, st1989analysis, yan2017statistical}, and Dunn's test \cite{elliott2011sas} is applied for post-hoc analysis instead of Tukey's test \cite{abdi2010tukey}. 


\vspace{0.3em}
\noindent \textbf{Confidence interval of the difference.} If a significant difference in means is observed, we further generate the confidence interval of the mean difference between groups  \cite{devore2012modern, hazra2017using} when the normality assumption is not violated.

%% file: sec/4_results.tex
\section{Experiments and Results}\label{Res-all}


Table \ref{tab:comparison} presents a comparative analysis of model performance. 
 The results demonstrate that all models achieve similar performance in terms of Dice scores and HD95 across all three labels, with U-Net marginally outperforming transformer-based models (Swin-UNETR and UNETR) and the Segresnet model.
 
\begin{table}[t]
  \centering
  \small
  \caption{Comparison of Dice Scores and HD95 Metrics for Different Models}
  \begin{tabular}{lcccccccc}
    \hline 
    \multirow{2}[4]{*}{Model} & \multicolumn{4}{c}{Dice Score [-]} & \multicolumn{4}{c}{HD95 [mm]} \\
    \cmidrule(lr){2-5} \cmidrule(lr){6-9}
    & NCR & ET & ED & Avg & NCR & ET & ED & Avg \\
    \hline
    U-Net & 70.33\% & 81.26\% & 84.79\% & 78.79\% & 6.99 & 5.10 & 4.56 & 5.55 \\ 
    Segresnet & 69.88\% & 80.30\% & 84.16\% & 78.11\% & 7.57 & 7.46 & 5.04 & 6.69 \\
    UNETR & 69.45\% & 80.55\% & 83.95\% & 77.98\% & 7.38 & 6.24 & 5.22 & 6.28 \\
    Swin-UNETR & 69.32\% & 81.29\% & 85.25\% & 78.62\% & 7.38 & 5.60 & 5.21 & 6.06 \\ 
    \hline
  \end{tabular}
  \label{tab:comparison}
\end{table}



Next, contrast-level Shapley values for each metric, averaged over three labels, are computed using four model architectures across five data folds. We define the matrix of contrast-level Shapley values for each combination of metric \( M \in \{\text{Dice, HD95} \} \), model \( \omega \in \{\text{U-Net, SegResNet, UNETR, Swin-UNETR} \} \), and fold \( f = 1, \dots, 5 \) as:  
\begin{equation}\label{Shap-0}
    \mathbf{\Phi}^{\omega,f}(M) = 
     \begin{pmatrix}
        \phi_{t1n,1}^{\omega, f}(M) & \phi_{t1n,2}^{\omega, f}(M) & \cdots & \phi_{t1n,J_f}^{\omega, f}(M) \\
        \phi_{t1c,1}^{\omega, f}(M) & \phi_{t1c,2}^{\omega, f}(M) & \cdots & \phi_{t1c,J_f}^{\omega, f}(M) \\
        \phi_{t2w,1}^{\omega, f}(M) & \phi_{t2w,2}^{\omega, f}(M) & \cdots & \phi_{t2w,J_f}^{\omega, f}(M) \\
        \phi_{t2f,1}^{\omega, f}(M) & \phi_{t2f,2}^{\omega, f}(M) & \cdots & \phi_{t2f,J_f}^{\omega, f}(M) \\
    \end{pmatrix}, \mathbf{\Phi}^{\omega,f}(M)  \in \mathbb{R}^{4,J_f},
\end{equation}
where \( \phi_{i,j}^{\omega, f}(M) \) represents the Shapley value for the \( j \)-th subject in fold \( f \), given contrast \( i \), model \( \omega \), and metric \( M \). We use \( J_f \) to denote the total number of subjects in fold \( f \).

For a given combination \((M, \omega, f)\), the contrast-wise vector \( \mathbf{C}_i^{\omega, f}(M) \) (\( i \in \{\text{t1n, t1c, t2w, t2f} \} \)) and subject-wise vector \( \mathbf{S}_j^{\omega, f}(M) \) (\( j = 1, \dots, J_f \)) are defined as follows:
\begin{equation}\label{shap-i}
\begin{array}{c}
    \mathbf{C}_i^{\omega, f}(M) = \mathbf{\Phi}_{i,\cdot}^{\omega, f}(M) = \left( \mathbf{\phi}_{i,1}^{\omega, f}(M), \ \mathbf{\phi}_{i,2}^{\omega, f}(M), \cdots, \ \mathbf{\phi}_{i,J_f}^{\omega, f}(M) \right), \mathbf{C}_i^{\omega, f}(M) \in \mathbb{R}^{J_f} \\
    \mathbf{S}_j^{\omega, f}(M) = \mathbf{\Phi}_{\cdot,j}^{\omega, f}(M) = \left( \mathbf{\phi}_{t1n,j}^{\omega, f}(M), \ \mathbf{\phi}_{t1c,j}^{\omega, f}(M), \mathbf{\phi}_{t2w,j}^{\omega, f}(M), \mathbf{\phi}_{t2f,j}^{\omega, f}(M)\right)^T, \mathbf{S}_j^{\omega, f}(M) \in \mathbb{R}^{4} 
\end{array}
\end{equation}  

In this study, we utilized four NVIDIA A40 GPUs to train our deep learning model and calculate the Shapley value. The evaluation time for each fold and model is approximately 1–2 minutes per subject.

\subsection{Shapley-based prediction insights: a clustering analysis}\label{res1} 

To analyze how segmentation performance overlaps with model weighting of MRI contrasts via contrast-level Shapley values, we applied k-means clustering. For each model-metric pair \((M, \omega)\), clustering was performed on the $\mathbf{S}_j^{\omega, f}(M)$ across five folds, i.e., \( \cup_{f=1}^{5} \cup_{j=1}^{J_f} \{ \mathbf{S}_j^{\omega, f}(M) \} \). 

\begin{figure}[htbp]
\centering
\subfigure[UNETR]{
    \includegraphics[width=0.4\linewidth, trim=0.2cm 0.5cm 0cm 0cm, clip]{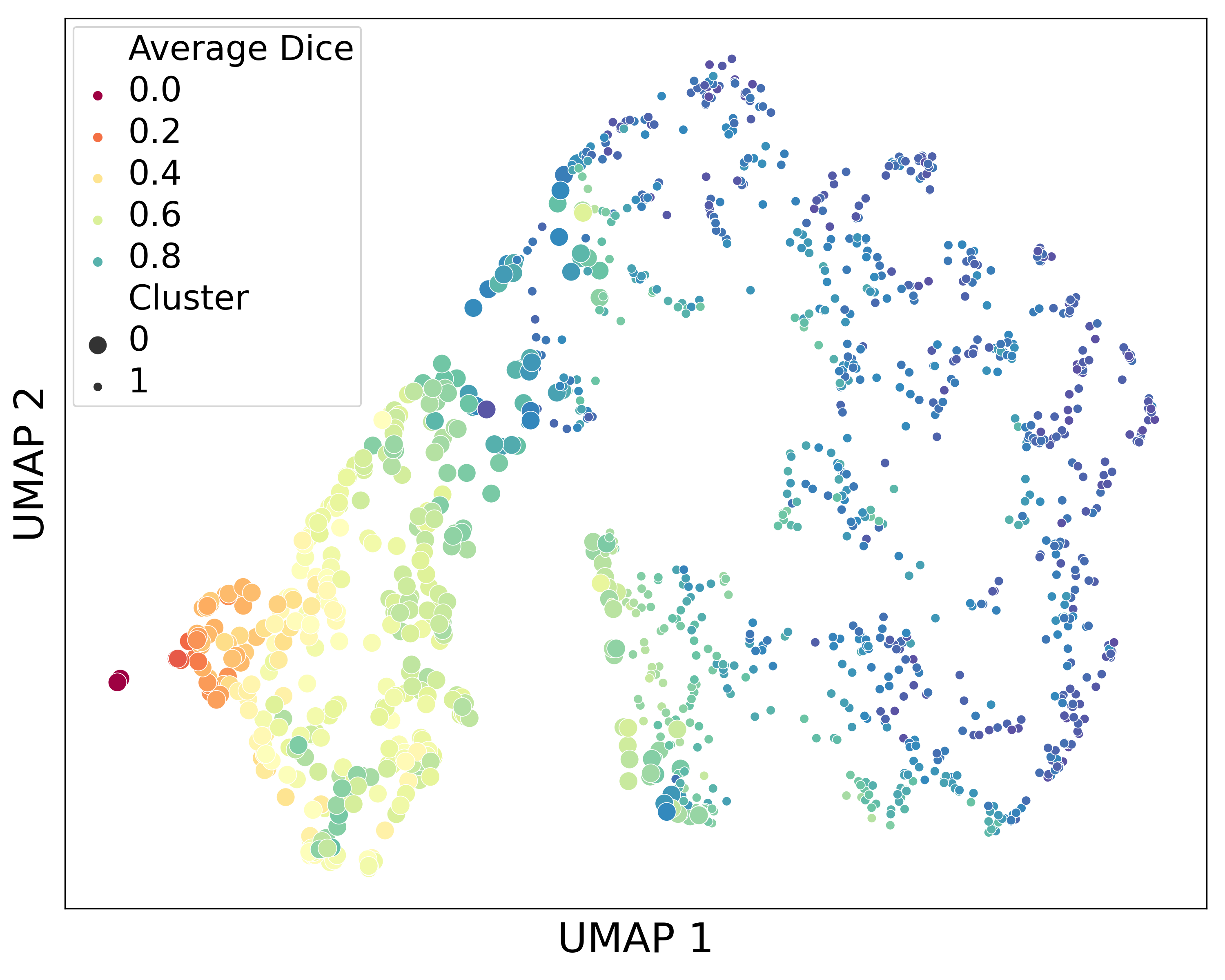}
    \label{fig:1a}
}
\hspace{0.05\linewidth}
\subfigure[SEGRES]{
    \includegraphics[width=0.4\linewidth, trim=0.2cm 0.5cm 0cm 0cm, clip]{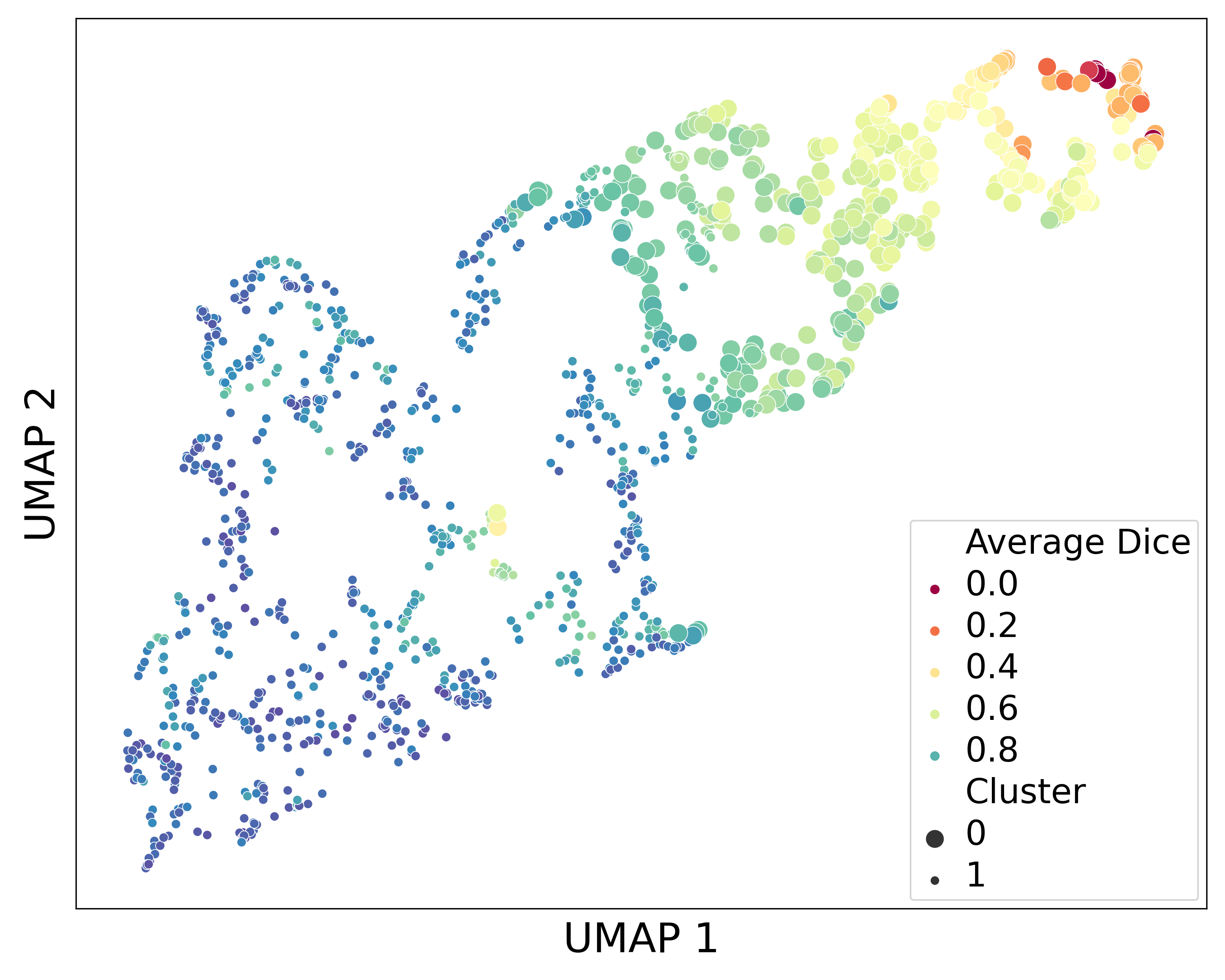}
    \label{fig:1b}
}
\caption{Clustering results on (a) {\scriptsize \( \cup_{f=1}^{5} \cup_{j=1}^{J_f} \{ \mathbf{S}_j^{\text{Unet}, f}(\text{Dice}) \} \) } and (b) {\scriptsize \( \cup_{f=1}^{5} \cup_{j=1}^{J_f} \{ \mathbf{S}_j^{\text{Swin-UNETR}, f}(\text{Dice}) \} \)} are visualized using UMAP for dimensionality reduction. The color represents the Dice score; the size of the dot is used to differentiate between cluster labels. }
\label{fig:1}
\end{figure}

We then use UMAP (Uniform Manifold Approximation and Projection) \cite{mcinnes1802umap} to visualize the clusters of Shapley value embeddings. Figure \ref{fig:1} illustrates an example with a significant pattern. For U-Net and Swin-UNETR, Shapley embedding clusters differentiate subjects with higher Dice scores from those with lower Dice scores.

\subsection{Shapley-based model prediction consistency: a comparative analysis    }\label{stat-res-all}



\subsubsection{Does each model learn consistent explanations? }\label{fold-res}

To assess the consistency of explanations across folds for each model, we analyzed the distribution of \( \mathbf{C}_i^{\omega, f}(M) \). The group standard deviation \( \sigma \) and mean \( \mu \) are key factors for determining distribution similarity, and statistical tests were applied to these metrics:
\begin{equation}\label{h0-1}
\begin{array}{l}
    H_0(\sigma | M, i, \omega): \sigma( \mathbf{C}_i^{\omega, 1}(M) ) = \sigma( \mathbf{C}_i^{\omega, 2}(M) ) = 
    \sigma( \mathbf{C}_i^{\omega, 3}(M) ) = 
    \sigma( \mathbf{C}_i^{\omega, 4}(M) ) = 
    \sigma( \mathbf{C}_i^{\omega, 5}(M)), \\
    H_0(\mu   | M, i, \omega): \mu( \mathbf{C}_i^{\omega, 1}(M)) = \mu( \mathbf{C}_i^{\omega, 2}(M) ) = 
    \mu( \mathbf{C}_i^{\omega, 3}(M) ) = 
    \mu( \mathbf{C}_i^{\omega, 4}(M) ) = 
    \mu( \mathbf{C}_i^{\omega, 5}(M)).
\end{array}
\end{equation}
If significant differences in mean or standard deviation are found, we conclude that inconsistent explanations are present across folds for a given pair of \( (M, i, \omega) \).

\begin{figure}[htbp]
\centering
\subfigure[U-net]{
\includegraphics[width=0.48\linewidth]{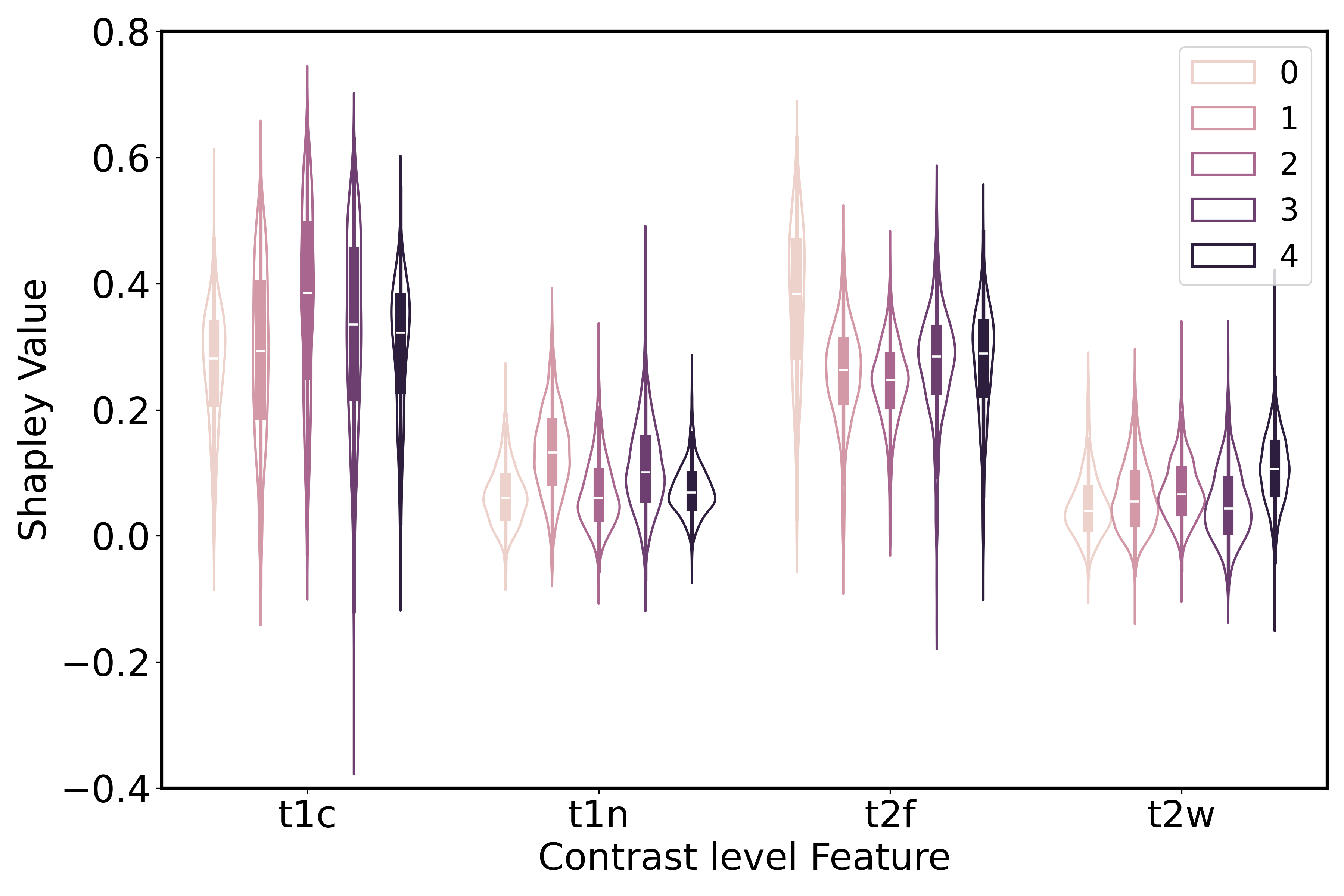}
    \label{fig:shapdista}
}\hfill
\subfigure[Swin-UNETR]{
\includegraphics[width=0.48\linewidth]{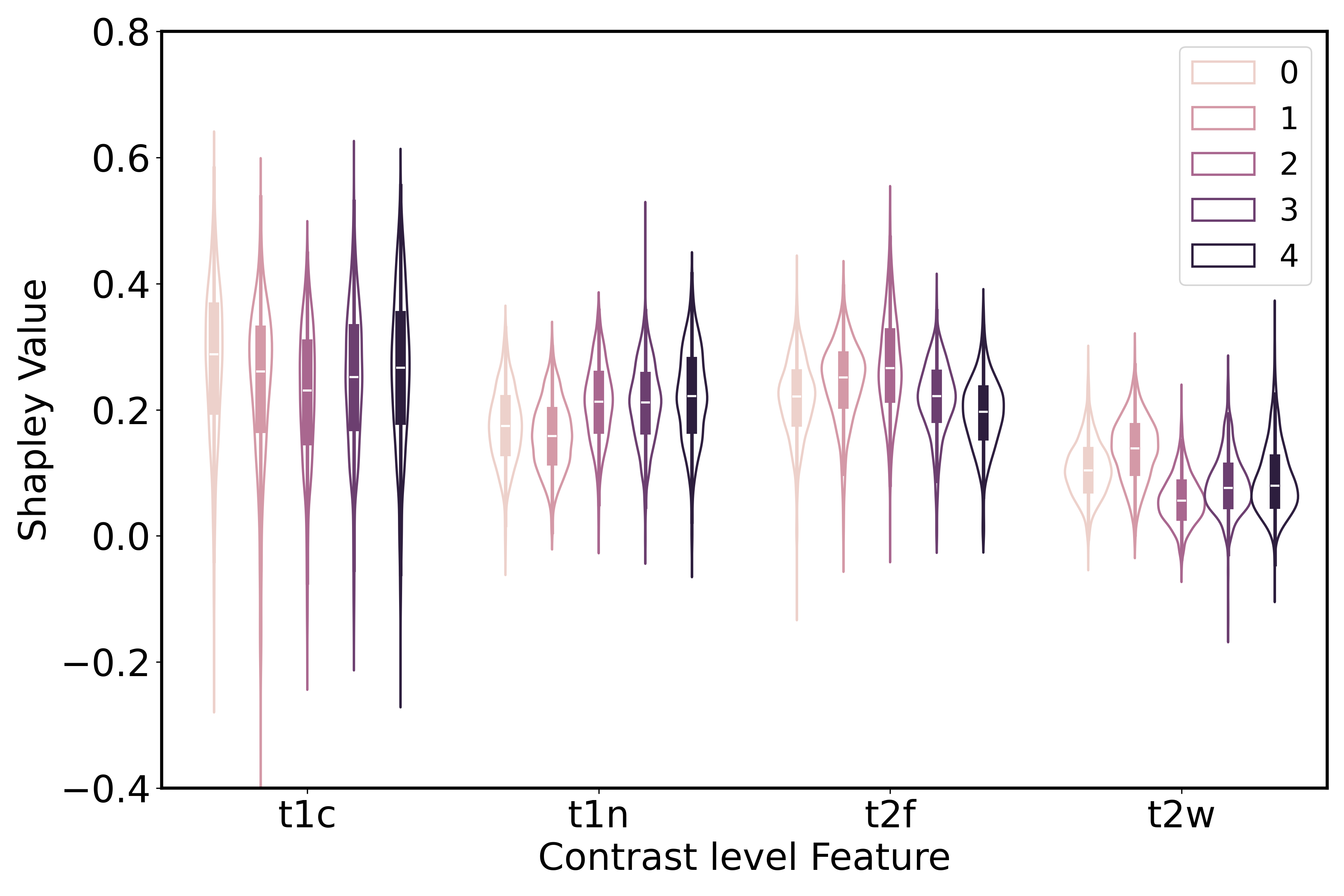}
    \label{fig:shapdistb}
}\\[1em]
\caption{The contrast-level Shapley values for all folds are computed based on the Dice score in each model. Panels (a) and (b) display the case of UNet and Swin-UNETR models, respectively.}
\label{fig:shapdist}
\end{figure}

Since the normality assumption for the Shapley value distribution \( \mathbf{C}_i^{\omega, f}(M) \) could not be guaranteed for some contrasts $i$, as indicated by the normality tests and non-zero skewness (Figure \ref{fig:shapdist}), Levene’s test, Kruskal-Wallis, and Dunn's post-hoc tests were applied. %

For all combinations of \((M, i, \omega)\), we get \( p < 0.01 \) in all 32 Levene's tests, rejecting \(H_0(\sigma|M, i, \omega)\) and indicating unequal variances across the five folds. Similarly, all 32 Kruskal-Wallis tests yield \( p < 0.01 \), rejecting \(H_0(\mu|M, i, \omega)\) and suggesting unequal means. These results invalidate the assumption that ``Model \(\omega\) learns consistent explanations across all five folds using contrast \(i\) for metric \(M\) evaluation," indicating significant differences in variance and means for at least one fold pair of each \((M, i, \omega)\) combination. 



Post-hoc tests are conducted to evaluate which pairs ($f_j, f_j'$) shows consistency explanation with the following null hypothesis:
\begin{equation}\label{h0-2}
\begin{array}{l}
    H_0(\mu | M, i, \omega, (f_j, f_{j'})) : \mu( \mathbf{C}_i^{\omega, f_j}(M)) = \mu( \mathbf{C}_i^{\omega, f_{j'}}(M)), f_j, f_{j'} \in \{1, 2, \cdots, 5\}; f_j \neq f_{j'}.
\end{array}
\end{equation}
Dunn's post-hoc tests reveal no significant differences in the \( t1c \) explanation between fold pairs 1 \& 5, 2 \& 3, 2 \& 4, and 4 \& 5 for Swin-UNETR, while significant differences exist in all other tests (Table \ref{tab:post-hoc1}). For example, in Table \ref{tab:post-hoc1}, \( p = 0.038 \) in the 1$^{\text{st}}$ column, the null hypothesis {\scriptsize $\mu( \mathbf{C}_{t1c}^{\text{\tiny Swin-UNETR}, 1}(Dice)) = \mu( \mathbf{C}_{t1c}^{\text{\tiny Swin-UNETR}, 5}(Dice))$} is not rejected, indicating ``Swin-UNETR learns consistent \( t1c \) contrast-level explanations between the 1st and 5th folds."


\begin{table}[t]
  \small
  \caption{Post-hoc tests reveal the pairs of folds where no statistical difference exists.}
  \begin{tabular}{ccccc}
    \hline 
    $H_0( { \tiny \mu | \text{Dice}, \text{t1c}, \text{SU}, \text{(1,5)} } )$ & 
    $H_0( { \tiny \mu | \text{Dice}, \text{t1c}, \text{SU}, \text{(2,3)} } )$ &
    $H_0( { \tiny \mu | \text{Dice}, \text{t1c}, \text{SU}, \text{(2,4)} } )$ &    
    $H_0( { \tiny \mu | \text{Dice}, \text{t1c}, \text{SU}, \text{(4,5)} } )$ &    All other tests \\
     \hline
     $p = 0.0385$ &$p = 0.0442$ &$p = 0.0687$ & $p = 0.0107$ & $p<0.01$\\ 
    \hline
  \end{tabular}

  {\scriptsize *Note that, we abbreviate Swin-UNETR as SU in this table.}
  \label{tab:post-hoc1}
\end{table}






\subsubsection{Do different models learn consistent explanations?}\label{model-res}

We first visualize the contrast-level Shapley value across all five folds for U-net, $\mathbf{C}_i^{\text{U-net},f}(\text{Dice})$, and Swin-UNETR, $\mathbf{C}_i^{\text{Swin-UNETR},f}(\text{Dice})$, using violin plot in Figure \ref{fig:shapdist}. We could observe that $t1c$ and $t2f$ are the most important image contrasts with the highest contrast-level Shapley value, this finding is consistent with the clinical explanation where $t2f$ suppresses cerebrospinal fluid signal, making edema and infiltration more visible, while $t1c$ provides clear delineation of enhancing tumor  (see section \ref{data}). We can also observe from this figure that Swin-UNETR weights $t1n$ significantly higher than U-Net.

To further investigate how model explanations are different within folds, we follow the procedure from Section \ref{fold-res}, with the key difference being that we compare results across multiple models while fixing the fold, unlike the previous tests where the models were fixed:
\begin{equation}
\begin{array}{c}
    H_0(\sigma | M, i, f): \sigma( \mathbf{C}_i^{\text{\tiny U-Net}, f}(M)) = \sigma(\mathbf{C}_i^{\text{\tiny Segresnet}, f}(M)) = \sigma(\mathbf{C}_i^{\text{\tiny UNETR}, f}(M)) = \sigma(\mathbf{C}_i^{\text{\tiny Swin-UNETR}, f}(M)), \\
    H_0(\mu | M, i, f): \mu( \mathbf{C}_i^{\text{\tiny U-Net}, f}(M)) = \mu(\mathbf{C}_i^{\text{\tiny Segresnet}, f}(M)) = \mu(\mathbf{C}_i^{\text{\tiny UNETR}, f}(M)) = \mu(\mathbf{C}_i^{\text{\tiny Swin-UNETR}, f}(M)).
\end{array}
\end{equation}
For all combinations of ($M, i, f$), the assumption that ``Within each fold $f$, all models learned consistent explanations when using contrast $i$ for metric $M$" is invalid [Levene’s test ($p<0.01$), Kruskal-Wallis test ($p<0.01$) for all tests]. However, the post-hoc tests do not reveal generalizable patterns across the models similar to the conclusion we presented in Table \ref{tab:post-hoc1}. To highlight performance differences, we provide the confidence intervals. 

Since the distributions of Shapley values are independent across models, and for each input \( j \), the differences between Shapley values, \( \mathbf{\phi}_{i,j}^{\omega, f}(M) - \mathbf{\phi}_{i,j}^{\omega', f}(M) \) (\(\omega \neq \omega'\)), passed the normality test, we further assess the difference between models by evaluating the confidence interval $CI_{\alpha}(\mu(\mathbf{C}_i^{(\omega,\omega'), f}(M)))$ given a desired level $\alpha$, where we define:
\begin{equation}\label{CI}
\begin{array}{c}
         \mathbf{C}_i^{(\omega,\omega'), f}(M) = \left( \mathbf{\phi}_{i,1}^{\omega, f}(M) - \mathbf{\phi}_{i,1}^{\omega', f}(M), \ \cdots, \ \mathbf{\phi}_{i,J_f}^{\omega, f}(M) - \mathbf{\phi}_{i,J_f}^{\omega', f}(M) \right)^T, \\
\end{array}
\end{equation}
with $J_f$ denoting the total number of subjects in fold $f$ from Definition (\ref{Shap-0}).

Here, we focus on the model difference in t1n, to test the hypothesis that Swin-UNETR has a higher contrast shapley value compared to other models, indicating a more balanced shapley value distribution and less basis toward t1c and t2f.
The confidence intervals for the mean difference in Shapley values (Swin-UNETR minus the other models) indicate a \textbf{significant positive difference} at a confidence level of 0.95, suggesting that Swin-UNETR places more attention on the \( t1n \) contrast (Figure \ref{fig:shapdist}).

\begin{table}[t]
  \centering
  \caption{Confidence Interval for Model Difference. The results indicate that Swin-UNETR exhibits significantly higher $t1n$ Shapley values compared to all other models for the Dice score at a 95\% confidence level.  }
  \begin{tabular}{cccccc}
        \hline
          & $f=1$   & $f=2$   & $f=3$   & $f=4$   & $f=5$ \\
          \hline
    $CI_{0.95}(\mu(\mathbf{C}_{t1n}^{\text{\tiny (SU, U)}, f}(\text{Dice})))$    & [0.11,0.12]  & [0.02,0.03] &  [0.14,0.15] &  [0.09,0.10] &  [0.15,0.16] \\
     $CI_{0.95}(\mu(\mathbf{C}_{t1n}^{\text{\tiny (SU, S)}, f}(\text{Dice})))$    & [0.05,0.06] &  [0.09,0.11]  & [0.06,0.07] &  [0.01,0.02] &  [0.07,0.08] \\
    $CI_{0.95}(\mu(\mathbf{C}_{t1n}^{\text{\tiny (SU, UR)}, f}(\text{Dice})))$    & [0.06,0.07]  & [0.00,0.01]  & [0.16,0.17]  & [0.03,0.06] &  [0.11,0.12] \\
    \hline
    \end{tabular}%
    
{\scriptsize *Note that, we abbreviate Swin-UNETR as SU, U-Net as U, SegResNet as S, and UNETR as UR in this table.}
  \label{tab:post-hoc1}
\end{table}

\begin{figure}[htbp]
\centering
\includegraphics[width=1.2\linewidth, trim= 7em 20em 3em 18em, clip]{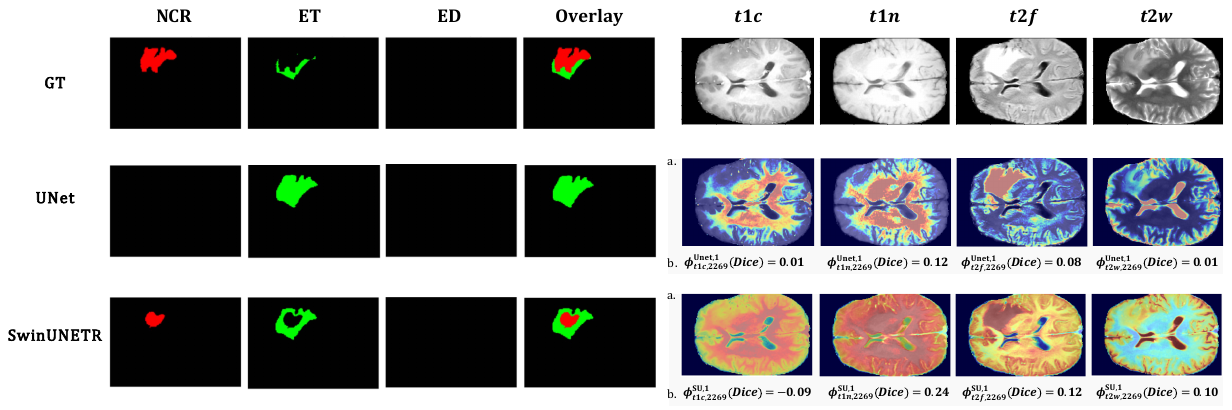}
\caption{Case comparison where Swin-UNETR outperforms U-Net. For the first four columns, from top to bottom, display: Ground truth, U-Net predictions, and Swin-UNETR predictions. For the last four columns, from top to bottom, display: input images, model explanations for U-Net (explanation (a) and (b)), and Swin-UNETR predictions (explanation (a) and (b)), where (a) shows GradCAM explanation for each contrast and (b) presents the proposed constrast-level Shapley values.}
\label{fig:3}
\end{figure}

To understand how transformer-based models differ from convolutional neural networks, we analyze cases where the Swin-UNETR model achieves a Dice score at least 20\% higher than U-Net and vice versa. Specifically, we examine cases where the Swin-UNETR model achieves a Dice score 25\% higher than U-Net (Figure \ref{fig:3}), and U-Net achieves a Dice score 23\% higher than Swin-UNETR (Figure \ref{fig:3}). This comparison highlights the advantages and limitations of each architecture in medical image segmentation tasks.



%% file: sec/5_Discussion.tex
\section{Discussion}
In this study, we systematically investigated the Shapley value for model explanation in multi-contrast medical image segmentation. 

Our proposed contrast-level Shapley explainability framework has three key contributions: (1) It is the first study to use Shapley analysis to explain multi-contrast medical image segmentation; (2) It is the first paper to analyze how different network structures weigh various MRI contrasts when making segmentation decisions; (3) It enhances clinical relevance by providing deeper insights into model performance with aggregate contributions of each MRI contrast in the tumor segmentation process, which is inherently interpretable by neuroradiologists, as they detect lesions by analyzing differences between different MRI contrasts in clinical practice.

Specifically, the contrast-level Shapley value reveals the (in)consistency of each model's explanations. The statistics indicate that Swin-UNETR is the most robust among all tested architectures. Despite being trained on different folds, Swin-UNETR consistently learns invariant representations across data subsets, whereas other models show variations in their explanations across folds (Table \ref{tab:comparison}).

Moreover, the contrast-level Shapley value provides insights on the differences among model architectures. As shown in Figure \ref{fig:shapdist}, the model explanations indicate that U-Net exhibits a bias toward features from $t1c$ and $t2f$, while Swin-UNETR distributes its explanations more evenly across contrasts. This was further confirmed by comparing $t1n$ Shapley values across different models, which revealed statistically higher Shapley values for Swin-UNETR (Table \ref{tab:post-hoc1}).


We also present a case in Figure \ref{fig:3} to demonstrate how explanations of different models could provide key insights into model failure. As discussed before, the training data includes 3 different tumor subtypes (see section \ref{data}). The innermost component of the tumor (shown in red in Figure \ref{fig:3}) is necrotic tissue in glioblastoma and meningioma, however, in metastasis, the definition of the innermost component is any tumor component that is not enhancing (but not necrotic). This implies that in $t2f$ images, the necrotic core will appear dark but non-enhancing metastatic tumor core and edema will appear bright.

Due to its dependence on contrasts with the highest intensity differences, namely $t1c$ and $t2f$, the U-Net architecture fails to accurately capture the innermost component (NCR). This suggests a potential bias towards $t1c$ and $t2f$, as indicated by the distribution of $ C_{t1c}^{\omega, f}(\text{Dice})$ and $ C_{t2f}^{\omega, f}(\text{Dice})$ exhibiting a significantly higher central tendency compared to $ C_{t1n}^{\omega, f}(\text{Dice})$ and $ C_{t2w}^{\omega, f}(\text{Dice})$ across all folds $f$ and models $\omega \in \{$UNET, Seg-Resnet, UNETR, Swin-UNETR 
$\}$, as shown in Figure 2 and supported by statistical tests in Section \ref{stat-res-all}. This bias may contribute to confusion with edema prediction, causing over-prediction relying on $t2f$ (edema appears bright as shown in Figure \ref{fig:3}). However, swin-UNETR effectively learns both local and global relationships within different contrasts through its self-attention mechanism, and was able to more accurately localize the tumor core in this challenging case.

Finally, for this case, we provide a comparison between GradCAM and our proposed contrast-level Shapley. As seen in Figure \ref{fig:3}, pixel-level explanations provided by GradCAM on each MRI contrast show model differences in terms of using pixel-level features. The heatmap of Swin-UNETR is more smooth while the heatmap of U-Net highlights only a few regions, but both of the explanations fail to capture clinically relevant explanations regarding contrast-level importance. For example, in Swin-UNETR, GradCAM exhibits a higher attention to $t1c$ compared to $t2f$. However, Contrast Shapley reveals that t1c negatively impacts the final Dice score, with a lower impact magnitude compared to $t2f$.

\section{Conclusion}
In this study, we propose Contrast Shapley for multi-contrast glioma segmentation. This method provides a quantitative framework for model explanation, offering insights into the fundamental characteristics of different deep learning architectures.